\documentclass[a4paper]{jpconf}

\usepackage{graphicx}
\begin{document}
\title{$^{31}$P NMR studies of an iron-based superconductor Ba$_{0.5}$Sr$_{0.5}$Fe$_{2}$(As$_{1-x}$P$_{x}$)$_{2}$ with $T_\mathrm{c}$ = 29 K}

\author{Yutaka Itoh$^1$, and Seiji Adachi$^2$}
\address{$^1$Department of Physics, Graduate School of Science, Kyoto Sangyo University, Kamigamo-Motoyama, Kika-ku, Kyoto 603-8555, Japan}    
\address{$^2$Superconducting Sensing Technology Research Association, 2-11-19 Minowa, Kohoku, Yokohama, Kanagawa 223-0051, Japan}

\ead{yitoh@cc.kyoto-su.ac.jp}

\begin{abstract}
We report $^{31}$P NMR studies of an oriented polycrystalline superconductor of Ba$_{0.5}$Sr$_{0.5}$Fe$_{2}$(As$_{1-x}$P$_{x}$)$_{2}$ with $x\sim$ 0.4 ($T_\mathrm{c}$ = 29 K) at $T$ = 14$-$325 K and $B_{0}$ = 1 and 5 T.  
The $^{31}$P Knight shift $K_{ab}$ at $B_{0}\perp c$ shows a nearly $T$-independent uniform spin susceptibility above $T_\mathrm{c}$ and a spin singlet decrease below $T_\mathrm{c}$. 
The $^{31}$P nuclear spin-lattice relaxation rate 1/$T_{1}$ shows an asymptotic behavior of $a+bT$ ($a$ and $b$ are constants) at $T >$ 100 K and the minimum at 40 K with an upturn toward $T_\mathrm{c}$. The $a$ term in 1/$T_{1}$ indicates the presence of two-dimensional antiferromagnetic spin fluctuations. The negative $\theta$ = $-$15 K of the Curie-Weiss-type antiferromagnetic spin susceptibility $\chi$($Q$) $\propto$ 1/($T$+$\theta$) in the analysis of 1/$T_{1}T$ suggests antiferromagnetic instability in the superconducting state.  
Discussions are made from the self-consistent renormalization (SCR) theory for the spin fluctuations with interlayer correlation. 
\end{abstract}

\section{Introduction}
The coexistence of superconductivity and antiferromagnetism has attracted great interests.
Although the superconducting transitions in weakly antiferromagnetic states (the superconducting transition temperature $T_\mathrm{c}$ $<$ the N$\acute\mathrm{e}$el temperature $T_\mathrm{N}$) have been observed widely in the underdoped superconducting compounds~\cite{R1,R2,R3}, the antiferromagnetic phase transitions in the superconducting states ($T_\mathrm{N} < T_\mathrm{c}$) remain to be obscure experimentally. 

An itinerant antiferromagnet BaFe$_2$As$_2$ (Ba122) has the N$\acute\mathrm{e}$el temperature $T_\mathrm{N}$ = 135 K~\cite{Ba122Kitagawa}. 
The high-$T_\mathrm{c}$ superconductivity with the optimal $T_\mathrm{c}\sim$ 30 K has been found in isovalent P substituted BaFe$_{2}$(As$_{1-x}$P$_{x}$)$_{2}$~\cite{Ba122}, Ba$_{0.5}$Sr$_{0.5}$Fe$_{2}$(As$_{1-x}$P$_{x}$)$_{2}$~\cite{Adachi,Tajima}, and SrFe$_{2}$(As$_{1-x}$P$_{x}$)$_2$~\cite{Tajima,Eisaki}.
In the underdoped regime with respect to the P concentration for BaFe$_{2}$(As$_{1-x}$P$_{x}$)$_{2}$, NMR measurements revealed that the superconductivity emerges in the weakly antiferromagnetic state at $T_\mathrm{c} < T_\mathrm{N}$~\cite{Ba122Iye}.
Systematic $^{31}$P NMR studies have been performed for BaFe$_{2}$(As$_{1-x}$P$_{x}$)$_{2}$~\cite{Ba122Nakai} and SrFe$_{2}$(As$_{1-x}$P$_{x}$)$_2$~\cite{Sr122PNMR}. 
The successive phase transitions from spin-density-wave (SDW) antiferromagnets to unconventional superconductors ($T_\mathrm{c} < T_\mathrm{N}$) have been studied with theoretical models~\cite{SDW1,SDW2}.

In this paper, we report $^{31}$P NMR measurements for an optimally doped superconductor Ba$_{0.5}$Sr$_{0.5}$Fe$_{2}$(As$_{1-x}$P$_{x}$)$_{2}$ with $x\sim$ 0.4 ($T_\mathrm{c}$ = 29 K). $^{31}$P nuclei can probe antiferromagnetic Fe spin fluctuations through off-diagonal hyperfine coupling constants~\cite{Ba122Kitagawa,Sr122Kitagawa,Shannon}.
We found the two-dimensional weakly antiferromagnetic spin susceptibility from the analysis of the $^{31}$P nuclear spin-lattice relaxation rate 1/$T_{1}$. 

\section{Experiments} 
Polycrystalline powder samples of Ba$_{0.5}$Sr$_{0.5}$Fe$_{2}$(As$_{1-x}$P$_{x}$)$_{2}$ with the nominal composition of $x$ = 0.4 ($T_\mathrm{c}$ = 29 K) were synthesized by a solid-state reaction method~\cite{Adachi}. 
The sample is an optimally doped or less optimally doped superconductor. 
The actual composition $x$ may be less than the nominal value of 0.4, because we observed a weak $^{31}$P NMR signal with no frequency shift associated with unreacted nonmagnetic phosphorous compounds.
The powder samples mixed in epoxy (Stycast 1266) were oriented and cured in a magnetic field of 5.0 T at room temperature.
NMR experiments were performed for the magnetically $ab$-axis aligned powder samples.

A phase-coherent-type pulsed spectrometer was utilized for the $^{31}$P NMR (nuclear spin $I$ = 1/2, the nuclear gyromagnetic ratio $^{31}\gamma_n$/2$\pi$ = 17.237 MHz/T) experiments at $B_{0}$ = 1.0 T ($T <$ 260 K) and 5.0 T ($T >$ 77 K). 
NaH$_2$PO$_4$ aqueous solution was served for the $^{31}$P NMR reference frequency with no shift.    
$^{31}$P NMR frequency spectra were obtained from the Fourier-transformed $^{31}$P nuclear spin-echoes.    
$^{31}$P nuclear spin-lattice relaxation curves were obtained by an inversion recovery technique. 
The spin-echo intensity $E$($t$) was recorded as a function of an interval time $t$ after an inversion pulse and $p$($t$) = $E$($\infty$) $-$ $E$($t$). 
The nuclear spin-lattice relaxation time $^{31}T_1$ was estimated by using the stretched exponential function with a variable exponent $\beta$ as 
\begin{eqnarray}
p(t) = p(0){\mathrm{exp}}\Bigl[- \Bigl({t \over {T_1}}\Bigr)^\beta\Bigr].  
\label{rec}
\end{eqnarray}
Irrespective of the variable exponent $\beta$, $T_1$ is a recovery time of the nuclear magnetization to $p(T_1)/p(0)$ = 1/$e$.   
The $\beta$ $<$ 1 indicates some of the nuclear spins with shorter relaxation times than $T_1$.
The results with the free-induction decays at $T >$ 77 K were consistent with those with the spin-echoes.   
\section{Experimental results}  
\subsection{$^{31}$P NMR spectra and Knight shifts}
\begin{figure}[t]
\begin{center}
\includegraphics[width=36pc]{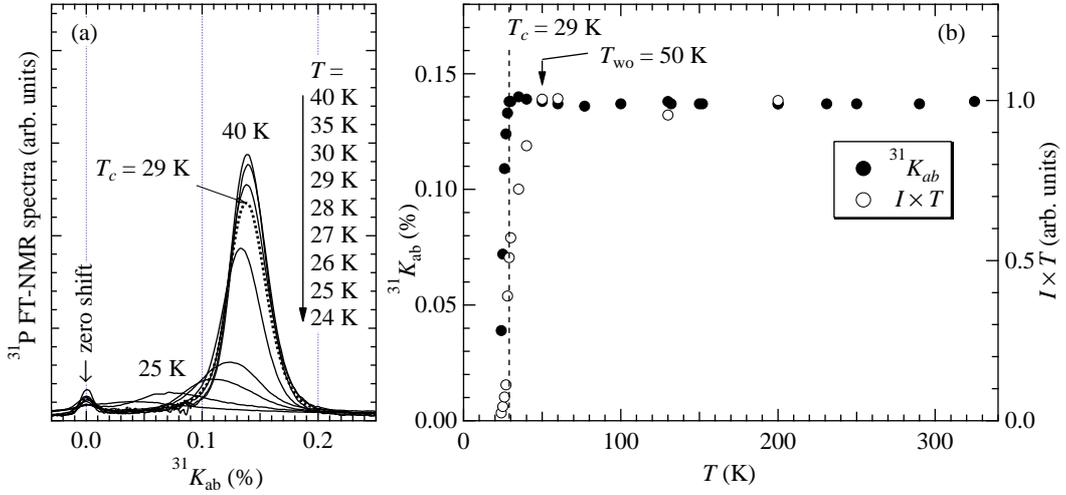}
\hspace{33pc}%
\vspace*{+0.5cm} 
\caption{\label{F1}
(a) Fourier-transformed $^{31}$P NMR spectra at $B_{0}\perp c$ and (b) Knight shifts $^{31}K_{ab}$ and integrated intensity multiplied by temperature $I\times T$ for Ba$_{0.5}$Sr$_{0.5}$Fe$_{2}$(As$_{1-x}$P$_{x}$)$_{2}$ with $x\sim$ 0.4. 
A weak NMR spectrum with zero shift is associated with unreacted nonmagnetic phosphorous compounds. 
}
\end{center}
\end{figure} 
\begin{figure}[h]
\begin{center}
\includegraphics[width=37pc]{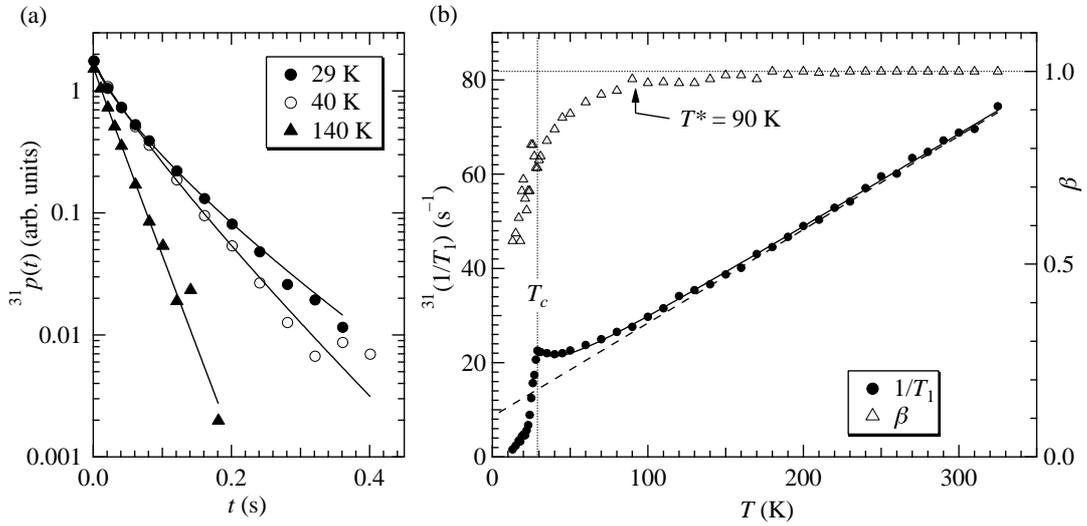}
\hspace{0pc}%
\caption{\label{F2}
(a) Recovery curves of $^{31}$P nuclear spin-echoes and (b) $^{31}$P nuclear spin-lattice relaxation rates $^{31}1/T_{1}$ and variable exponents $\beta$ for Ba$_{0.5}$Sr$_{0.5}$Fe$_{2}$(As$_{1-x}$P$_{x}$)$_{2}$ with $x\sim$ 0.4 at $B_{0}\perp c$.  
The solid curves in (a) are the least-squares fitting results using Eq.~(\ref{rec}). 
The variable exponent $\beta$ in (b) shows a decrease below $T^*$ = 90 K.
The solid curve in (b) is the least-squares fitting result using Eq.~(\ref{T1}). 
The dashed line in (b) is $a$ + $bT$ in an asymptotic behavior.
}
\end{center}
\end{figure} 
Figure~\ref{F1} shows the Fourier-transformed $^{31}$P NMR spectra (a) and the Knight shifts $^{31}K_{ab}$ and integrated intensity multiplied by temperature $I\times T$ (b). 
The integrated intensity of the NMR spectrum $I$ should increase with decreasing temperature in a Curie law and $I\times T$ should be independent of temperature. 
However, $I\times T$ decreases abruptly below $T_\mathrm{wo}$ = 50 K, which is a wipeout effect on NMR intesnity. 
Since the wipeout effect below $T_\mathrm{wo}$ = 50 K is not due to superconducting shielding effects, the loss of the NMR spectrum results from the short $T_2$ signals due to the emergence of slow fluctuations.  
Similar wipeout effects have been observed for BaFe$_2$(As$_{1-x}$P$_x$)$_2$~\cite{1storder}. 

$^{31}K_{ab}$ is the sum of the spin shift $K_\mathrm{spin}$ and the chemical shift $K_\mathrm{chem}$, that is $^{31}K_{ab}$ = $K_\mathrm{spin}$ + $K_\mathrm{chem}$. 
The chemical shift $K_\mathrm{chem}$ is estimated to be 0.018 $\%$ for BaFe$_{2}$(As$_{1-x}$P$_{x}$)$_{2}$~\cite{Ba122Nakai} and 0.04 $\%$ for SrFe$_{2}$(As$_{1-x}$P$_{x}$)$_{2}$~\cite{Sr122PNMR}. 
Then, $K_\mathrm{spin}$ is estimated to be 0.10 or 0.12 $\%$.
Figure~\ref{F1}(b) shows that the spin Knight shift $K_\mathrm{spin}$ is nearly independent of temperature above $T_\mathrm{c}$ = 29 K. 
Since the spin Knight shift is proportional to the uniform spin susceptibility, 
the uniform spin susceptibility is found to be nearly independent of temperature in the normal state.
No pseudogap effect characterizes the uniform spin susceptibility.

Figure~\ref{F1}(b) also shows that $K_\mathrm{spin}$ rapidly decreases on cooling below $T_\mathrm{c}$ = 29 K.
Since $^{31}K_{ab}$ is close to 0.04 at 24 K, the spin susceptibility tends to vanish in the superconducting state. 
Thus, the Cooper pairs are found in the spin singlet states.    

\subsection{$^{31}$P nuclear spin-lattice relaxation rate 1/$T_1$} 　
Figure~\ref{F2} shows the recovery curves of $^{31}$P nuclear spin-echoes (a) and $^{31}$P nuclear spin-lattice relaxation rates $^{31}(1/T_1)$ and variable exponents $\beta$ as functions of temperature (b).  
The variable exponent $\beta$ shows a decrease on cooling below $T^*$ = 90 K. 
The small exponent $\beta <$ 1 indicates the deviation in the recovery curve from a single exponential function to a stretched exponential function because of a distribution in the NMR relaxation time.
Similar behavior of $\beta$ has been observed for BaFe$_2$(As$_{1-x}$P$_x$)$_2$~\cite{Curro}. 
The $\beta \sim$ 0.8 at $T_\mathrm{c}$ tells a small distribution of the time constant.
The small distribution in the NMR relaxation time below $T^*$ may be associated with the onset of a nematic orbital ordering in BaFe$_2$(As$_{1-x}$P$_x$)$_2$~\cite{nematic}. 

In Fig.~\ref{F2}(b), the solid curve is the least-squares fitting result using the following equation 
\begin{eqnarray}
^{31}\Bigl({1 \over {T_1}}\Bigr) = T\Bigl({a \over {T + \theta}} + b\Bigr),  
\label{T1}
\end{eqnarray} 
where $a$ and $b$ are fitting parameters. The $a$ term is due to the antiferromagnetic spin fluctuations enhanced at a finite wave vector $Q$ and the $b$ term is due to the uniform spin fluctuations at and around $q$ = 0.
Since the uniform spin susceptibility [$\propto K_\mathrm{spin}$ in Fig.~\ref{F1}(b)] is nearly independent of temperature, the $b$ term is the Korringa process.
In Fig.~\ref{F2}(b), the dashed line is $a$ + $bT$ in an asymptotic behavior, where $a$ and $b$ are the fitting results using Eq.~(\ref{T1}).  
One should note the similar behaviors of 1/$T_1$ = $a$ + $bT$ for the planar Cu nuclei in the high-$T_\mathrm{c}$ cuprate superconductors~\cite{Imai0,Imai}. 
In contrast to the spin pseudogap effects on the underdoped cuprates, $^{31}(1/T_{1})$ takes the minimum value at about 40 K and 
increases toward $T_\mathrm{c}$. 
\begin{figure}[b]
\begin{center}
\includegraphics[width=36pc]{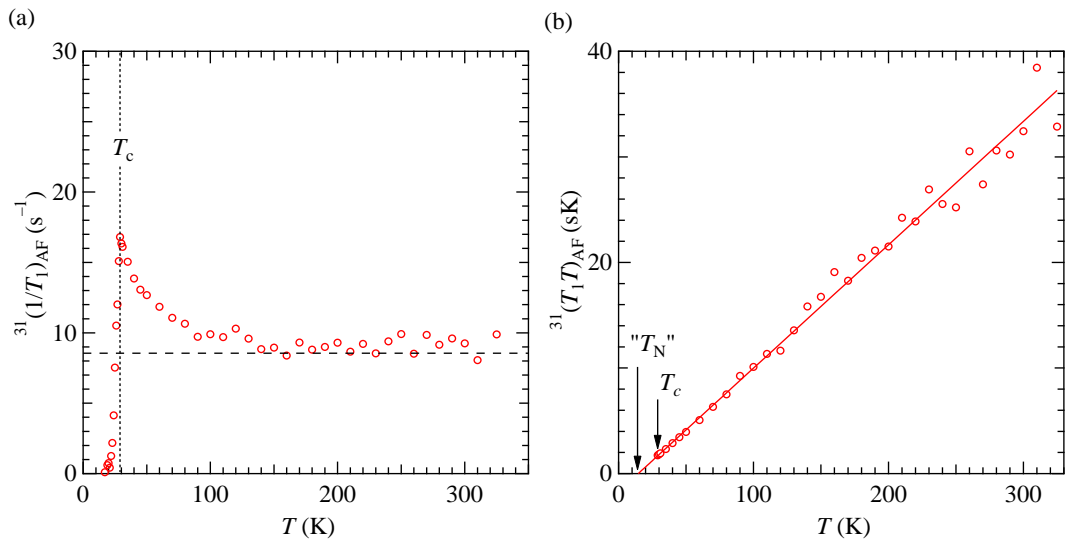}\hspace{0pc}%
\caption{\label{F3}
(a) $^{31}(1/T_1)_\mathrm{AF}$ and (b) $^{31}(T_1T)_\mathrm{AF}$ as functions of temperature for Ba$_{0.5}$Sr$_{0.5}$Fe$_{2}$(As$_{1-x}$P$_{x}$)$_{2}$ with $x\sim$ 0.4. 
The dashed line in (a) is $^{31}(1/T_1)_\mathrm{AF}$ $\rightarrow$ 8.6 s$^{-1}$. 
The solid line in (b) is proportional to ($T$ + $\theta$) with $\theta$ = $-$15 K.  
The inverse Curie-Weiss law well reproduces $^{31}(T_1T)_\mathrm{AF}$ from $T_\mathrm{c}$ to 325 K.
}
\end{center}
\end{figure} 
\begin{figure}[b]
\begin{minipage}{18pc}
\includegraphics[width=18pc]{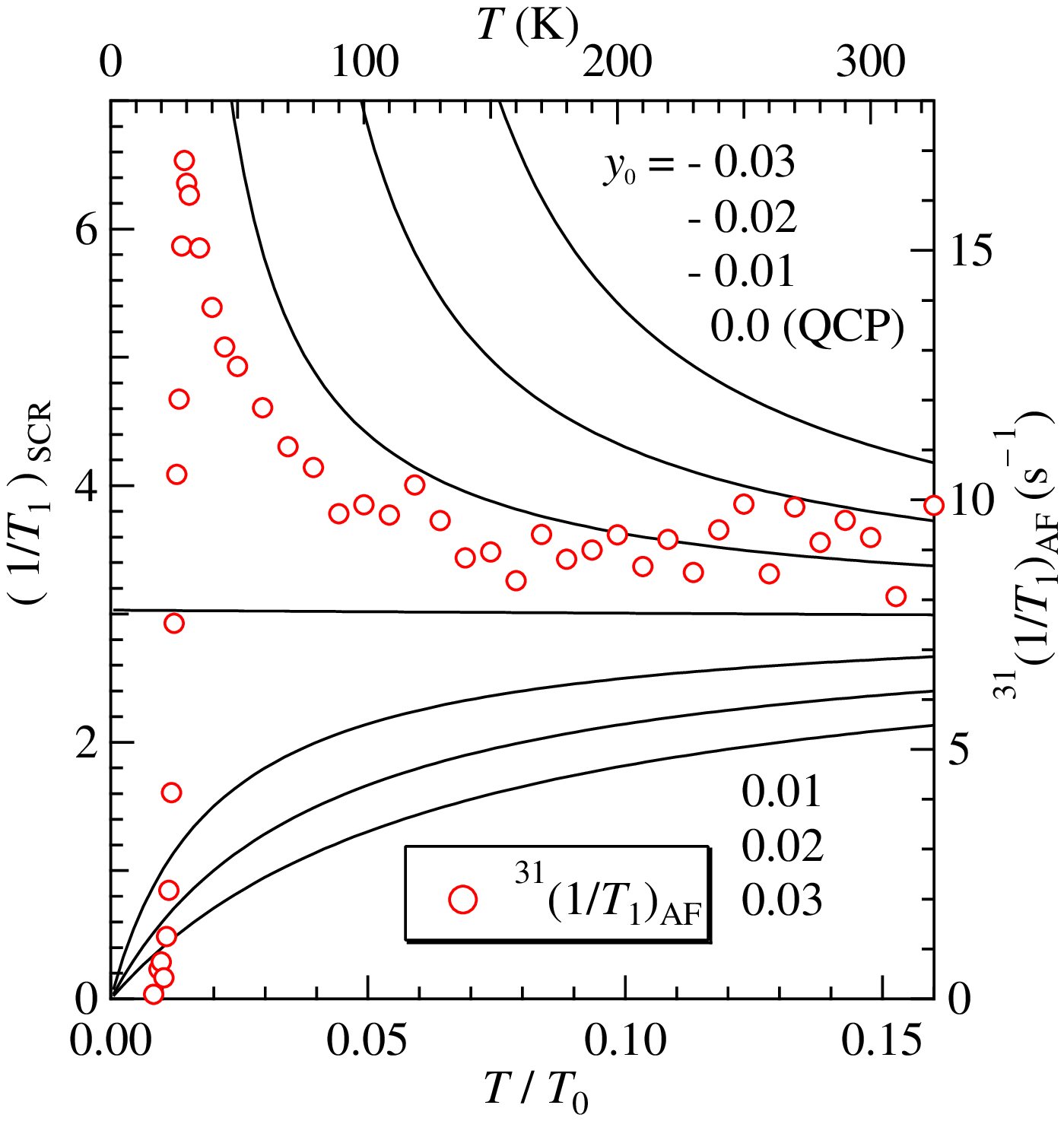}
\caption{\label{F4}
Solid curves are the theoretical (1/$T_1$)$_\mathrm{SCR}$'s plotted against a reduced temperature $T/T_0$ with the SCR theory~\cite{MTU}. 
Open circles are the experimental $^{31}(1/T_1)_\mathrm{AF}$ plotted against $T$ for Ba$_{0.5}$Sr$_{0.5}$Fe$_{2}$(As$_{1-x}$P$_{x}$)$_{2}$ with $x\sim$ 0.4.
}
\end{minipage}\hspace{2pc}%
\begin{minipage}{18pc}
\includegraphics[width=18pc]{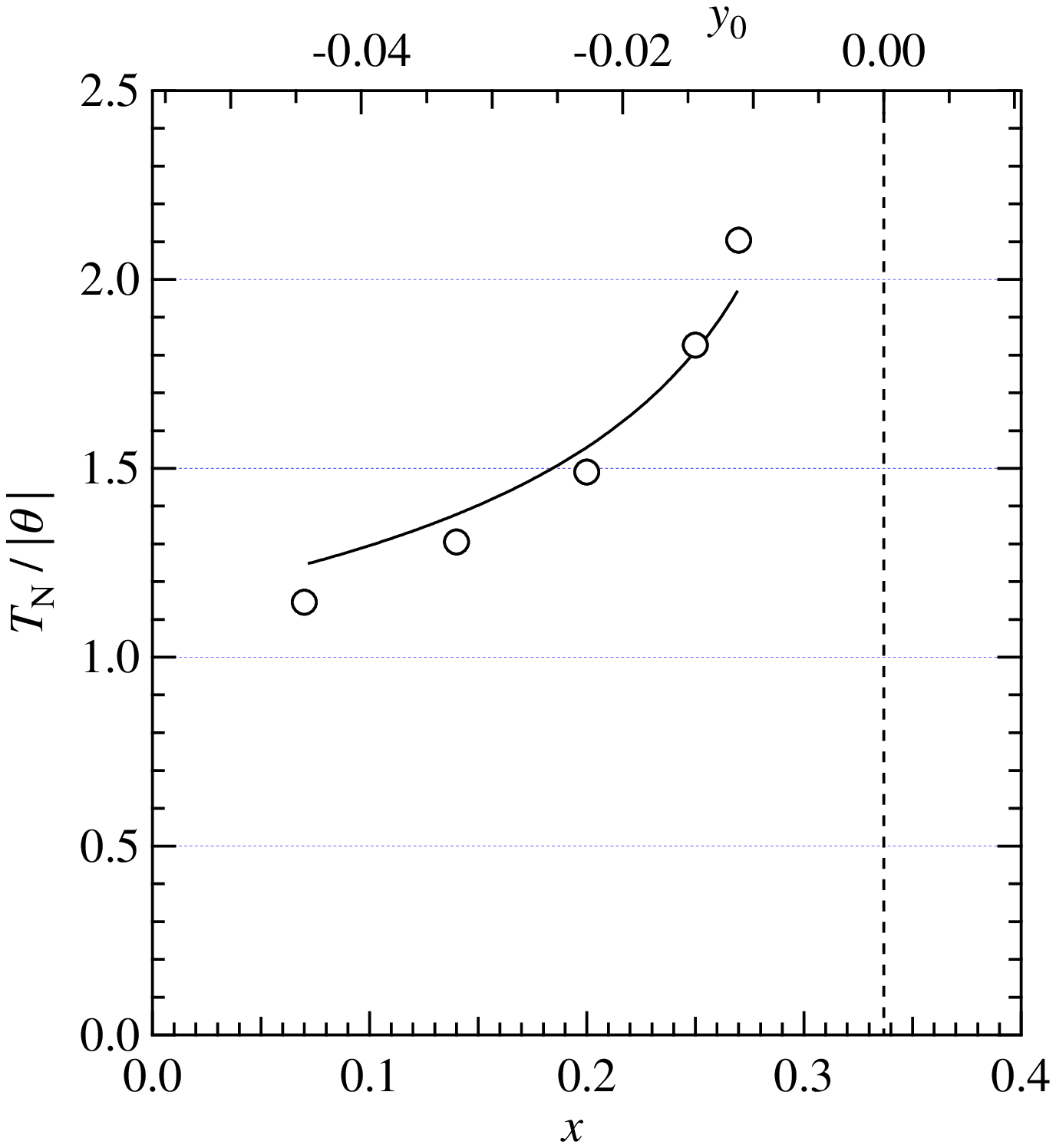}
\caption{\label{F5}
Open circles are $T_\mathrm{N}$/$|\theta|$ vs $x$ and $y_0$ in the SDW states of BaFe$_{2}$(As$_{1-x}$P$_{x}$)$_{2}$~\cite{Ba122TN}. 
The values $y_0$'s are our estimations.
The solid curve is the theoretical $T_\mathrm{N}$/$|\theta|$ vs $y_0$ of Eq.~(\ref{TNQ}). 
$T_\mathrm{N}$/$|\theta|\propto y_0^{-1/3}$ and $T_{A1}/T_A$ = 0.12 reproduce the experimental values. 
}
\end{minipage} 
\end{figure}
\section{Discussions}
\subsection{Weakly antiferromagnetic spin fluctuations}
We extracted the component of the relaxation rate due to the antiferromagnetic spin fluctuations $^{31}$(1/$T_1$)$_\mathrm{AF}$ from $^{31}$(1/$T_1)$ as
\begin{eqnarray}
^{31}\Bigl({1 \over {T_1}}\Bigr)_\mathrm{AF} \equiv  ^{31}\Bigl({1 \over {T_1}}\Bigr) - bT.  
\label{T1AF}
\end{eqnarray} 

Figure~\ref{F3} shows $^{31}(1/T_1)_\mathrm{AF}$ (a) and $^{31}(T_1T)_\mathrm{AF}$ (b) as functions of temperature. 
$^{31}(T_1T)_\mathrm{AF}$ is associated with the inverse staggered spin susceptibility 1/$\chi (Q)$.
The solid line in (b) is proportional to ($T$ + $\theta$) with $\theta$ = $-$15 K. 
``$T_\mathrm{N}$'' is a hypothetical N$\acute\mathrm{e}$el temperature given by $|\theta |$.
One should note that not only $^{31}(1/T_1T)_\mathrm{AF}$ but also $^{31}(1/T_1)_\mathrm{AF}$ shows the increase toward $|\theta |$ = 15 K.
$^{31}(1/T_1)_\mathrm{AF}$ takes a peak at $T_\mathrm{c}$. 

The self-consistent renormalization (SCR) theory for two-dimensional antiferromagnetic spin fluctuations tells us an approximate relation~\cite{MTU,ItohSCR} 
\begin{eqnarray}
(T_1T)_\mathrm{AF}\propto (2\alpha_s T_A)(T + \theta)
\label{T1SCR1}
\end{eqnarray}
and
\begin{eqnarray}
\theta = c_{\theta}y_0T_0  
\label{T1SCR2}
\end{eqnarray}
with $c_{\theta}\approx$ 1.22 for $-$0.04 $< y_0 <$ 0.01 (0.02 $< T/T_0 <$ 0.3 for $y_0$ = $-$0.01), and a mode-mode coupling parameter $y_1$ = 3. 
$y_0$ measures a distance from the quantum critical point (QCP). 
$T_A$ specifies the in-plane spatial spread of the spin fluctuations. 
The spin fluctuation energy $T_0$ specifies the frequency spread of the spin fluctuations.
$\alpha_s$ is associated with the Stoner enhancement factor with $\alpha_s$ = 2$I\chi_0(Q)$, where $\chi_0(Q)$ is a staggered spin susceptibility without an electron-electron interaction $I$.    
$y_0 < $ 0 indicates a weakly antiferromagnetic ground state, and $y_0 > $ 0 indicates a nearly antiferromagnetic ground state~\cite{MTU}. 

The negative Weiss temperature $\theta$ = $-$15 K indicates $y_0 < $ 0 for the weakly antiferromagnetic ground state. 
We estimated $y_0$ = $-$0.0061 using Eq.~(\ref{T1SCR2}) with $T_0$ = 2000 K~\cite{IND}.
Figure~\ref{F4} shows the theoretical (1/$T_1$)$_\mathrm{SCR}$'s plotted against a reduced temperature $T/T_0$ (the solid curves) from the SCR theory with several $y_0$'s~\cite{MTU} and the experimental $^{31}(1/T_1)_\mathrm{AF}$ plotted against $T$ (the open circles).
The SCR theory for the two-dimensional antiferromagnetic spin fluctuations with $y_0 < $ 0 reproduces the experimental $^{31}(1/T_1)_\mathrm{AF}$.

\subsection{Estimation of $T_\mathrm{N}$}
No finite temperature long-range ordering is a consequence of  the SCR theory for pure two-dimensional systems~\cite{MTU}.
An interlayer correlation can cause a finite temperature antiferromagnetic phase transition~\cite{Konno}. 
We obtain an approximate expression of a finite $T_\mathrm{N}$ for the spin fluctuations with the $c$ axis spatial extension $T_{A1}$ as 
\begin{eqnarray}
T_\mathrm{N} = c_N\Biggl({y_{0}^2 \over {y_{1}^2}}{T_{A1} \over {T_{A}}}\Biggr)^{1/3}T_0  
\label{TN}
\end{eqnarray}
with $c_N\approx$ 2.25 adapt from Refs.~\cite{Konno,Hasegawa,Nakayama}.
For BaFe$_{2}$(As$_{1-x}$P$_{x}$)$_{2}$, the $k_z$ dispersion of the Fermi surface~\cite{ARPES} and  
the three-dimensional character of the dynamical spin susceptibility~\cite{ND} enable us to estimate $T_{A1}/T_{A}\sim$ 0.1.
Using Eq.~(\ref{TN}) with $T_{A1}/T_A$ = 0.10, $y_0$ = $-$0.0061, $y_1$ = 3 and $T_0$ = 2000 K, we obtained $T_\mathrm{N}$ = 27 K for Ba$_{0.5}$Sr$_{0.5}$Fe$_{2}$(As$_{1-x}$P$_{x}$)$_{2}$ with $x\sim$ 0.4. 

Figure~\ref{F5} shows the experimental ratios of $T_\mathrm{N}$/$|\theta|$ vs $x$ and $y_0$ in the actual SDW states of BaFe$_{2}$(As$_{1-x}$P$_{x}$)$_{2}$ taken from Ref.~\cite{Ba122TN}.
Here, the values $y_0$'s were newly estimated by Eqs.~(\ref{T1SCR2}) and~(\ref{TN}). 
We believe that $T_\mathrm{N}\neq|\theta|$ is significant.
The sudden disappearance of $T_\mathrm{N}$ around the optimal superconductivity is referred to as a weakly first-order-like transition (an avoided QCP)~\cite{R3,1storder}, while the continuous diminishment of $|\theta|$ is referred to as a QCP~\cite{Ba122SCR}. 
$T_\mathrm{N}$ is the three-dimensional critical temperature, while $|\theta|$ is the two-dimensional characteristic temperature.
The solid curve in Fig.~\ref{F5} is the theoretical function
\begin{eqnarray}
{T_\mathrm{N}\over{|\theta}|} = {0.885 \over {|y_{0}|^{1/3}}}\Biggl({T_{A1} \over {T_{A}}}\Biggr)^{1/3} 
\label{TNQ}
\end{eqnarray}
with $y_1$ = 3 and a fitting parameter $T_{A1}/T_A$ = 0.12. 
The theoretical $y_0$ dependence of Eq.~(\ref{TNQ}) reproduces the experimental $x$ dependence of $T_\mathrm{N}$/$|\theta|$ in the SDW states.
Thus, the three dimensionality on $T_{A1}$ is a key in $T_\mathrm{N}\neq|\theta|$, and only $y_0$ close to zero yields $T_\mathrm{N} > |\theta|$.

The wipeout effect on the NMR spectra below $T_\mathrm{wo}$ = 50 K results from the development of unobservable NMR signals with short $T_2$ decay in Ba$_{0.5}$Sr$_{0.5}$Fe$_{2}$(As$_{1-x}$P$_{x}$)$_{2}$ with $x\sim$ 0.4, which suggests the emergence of slowly fluctuating local fields.
Although the wipeout effect is not direct evidence for a static N$\acute\mathrm{e}$el order, the neutron scattering studies indicate that the BaFe$_{2}$(As$_{1-x}$P$_{x}$)$_{2}$ superconductors with the wipeout effects exhibit the static N$\acute\mathrm{e}$el order below $T_\mathrm{wo}$~\cite{1storder}.
Thus, we speculate that the weakly antiferromagnetic phase transition at $T_\mathrm{N}$ = 27 K $<$ $T_\mathrm{c}$ may occur in the suppression of the superconductivity for Ba$_{0.5}$Sr$_{0.5}$Fe$_{2}$(As$_{1-x}$P$_{x}$)$_{2}$ with $x\sim$ 0.4. 
The coexistence of a SDW ordering in a nontrivial superconducting state is also theoretically possible~\cite{Machida}.     

\subsection{Gap parameter}
\begin{figure}[h]
\begin{center}
\includegraphics[width=18pc]{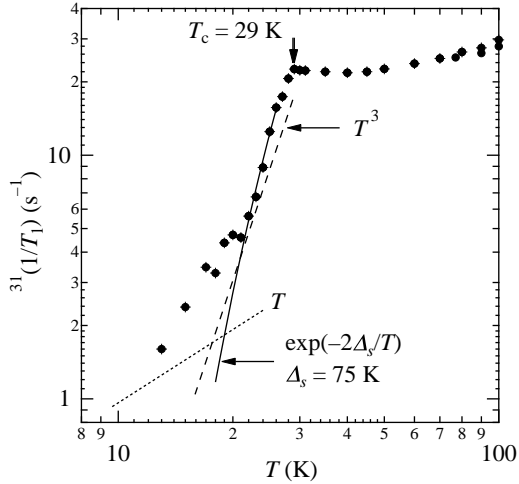}
\caption{\label{F6}
Log-log plot of $^{31}(1/T_1)$ of the observable NMR signal as a function of temperature (closed circles) for Ba$_{0.5}$Sr$_{0.5}$Fe$_{2}$(As$_{1-x}$P$_{x}$)$_{2}$ with $x\sim$ 0.4. 
$^{31}(1/T_1)$ below 24 K may be extrinsic.
} 
\end{center}  
\end{figure}
Figure~\ref{F6} shows the log-log plot of $^{31}(1/T_1)$ of the observable NMR signal as a function of temperature.
$^{31}(1/T_1)$ below 24 K may be extrinsic due to the NMR signal with zero shift in Fig.~\ref{F1}(a).
No Hebel-Slichter peak in $^{31}(1/T_1)$ is found just below $T_\mathrm{c}$ = 29 K.
The solid curve, dashed line, and dotted line are an activation function of $\mathrm{exp}(-2{\it\Delta_s}/T)$ with ${\it\Delta_s}$ = 75 K, a $T^3$ function in $1/T_1$ due to a $d$-wave gap parameter, and a $T$-linear function, respectively.
The activation function of $\mathrm{exp}(-2{\it\Delta_s}/T)$ reproduces a part of the $T$-dependence of $^{31}(1/T_1)$.
This is consistent with an extended $s$-wave ($s_{+-}$) gap structure of the superconducting order parameter~\cite{OP}.
The large ratio of 2${\it\Delta_s}/T_\mathrm{c}$ = 5.2 indicates a strong coupling superconductivity.
No information at lower temperatures is available because of the wipeout effect.

\subsection{Sr substitution for Ba}
One may expect two effects of the Sr substitution for Ba in BaFe$_{2}$As$_{2}$.
One is the chemical pressure effect.
The other is the randomness effect of crystalline potentials.
Although the superconductivity is observed in BaFe$_{2}$(As$_{1-x}$P$_{x}$)$_{2}$~\cite{Ba122} and physically pressed BaFe$_{2}$As$_{2}$~\cite{Press1,Press2},
no superconductivity is observed in isovalnet Sr substituted Ba$_{1-x}$Sr$_{x}$Fe$_{2}$As$_{2}$~\cite{NSC}.
The reason why no superconductivity emerges in Ba$_{1-x}$Sr$_{x}$Fe$_{2}$As$_{2}$ is attributed to the lack of shrinkage of the Fe-As bond length~\cite{WhyP}.
Since $T_\mathrm{N}$ increases from 135 K (BaFe$_{2}$As$_{2}$) to 199 K (SrFe$_{2}$As$_{2}$)~\cite{Sr122Kitagawa,WhyP}, the Sr substitution may enhance an interlayer coupling.
In passing, BaFe$_{2}$(As$_{1-x}$P$_{x}$)$_{2}$, Ba$_{0.5}$Sr$_{0.5}$Fe$_{2}$(As$_{1-x}$P$_{x}$)$_{2}$, and SrFe$_{2}$(As$_{1-x}$P$_{x}$)$_2$ have in common the optimal $T_\mathrm{c}$ = 30$-$33 K~\cite{Tajima}.
No remarkable effect of the random potentials was found in the $^{31}$P NMR spectra and relaxation rates for the present Ba$_{0.5}$Sr$_{0.5}$Fe$_{2}$(As$_{1-x}$P$_{x}$)$_{2}$ with $x\sim$ 0.4.

\section{Conclusions}
In conclusion, we observed the two-dimensional antiferromagnetic spin susceptibility $\chi$($Q$) $\propto$ 1/($T + \theta$) ($\theta = -$15 K) and the wipeout effect on the $^{31}$P NMR spectra below $T_\mathrm{wo}$ = 50 K for an iron-based superconductor Ba$_{0.5}$Sr$_{0.5}$Fe$_{2}$(As$_{1-x}$P$_{x}$)$_{2}$ with $x\sim$ 0.4 and $T_\mathrm{c}$ = 29 K.
We estimated a finite $T_\mathrm{N}$ = 27 K from the quasi two-dimensional SCR theory with $c$ axis correlation.
The $^{31}$P Knight shift $K_{ab}$ shows the nearly $T$-independent uniform spin susceptibility above $T_\mathrm{c}$ and
the spin singlet formation below $T_\mathrm{c}$.

\ack 
We thank  K. Ishida for fruitful discussions with hyperfine couplings. 

\section*{References}

\smallskip
 

\end{document}